\documentclass[aps,pre,twocolumn,groupedaddress]{revtex4-1}

\usepackage{amsmath,amssymb}
\usepackage{color}
\usepackage{graphicx}
\usepackage{bm}

\begin{document}


\title{Nonlinear Kalman filter based on duality relations between continuous and discrete-state stochastic processes}


\author{Jun Ohkubo}
\affiliation{
Graduate School of Science and Engineering, Saitama University,
255 Shimo-Okubo, Sakura, Saitama, 338-8570, Japan
}


\date{\today}

\begin{abstract}
A new application of duality relations of stochastic processes
is demonstrated.
Although conventional usages of the duality relations
need analytical solutions for the dual processes,
we here employ numerical solutions of the dual processes
and investigate the usefulness.
As a demonstration, estimation problems of hidden variables in stochastic differential
equations are discussed.
Employing algebraic probability theory,
a little complicated birth-death process is derived from the stochastic 
differential equations, and
an estimation method based on the ensemble Kalman filter is proposed.
As a result, the possibility for making faster computational algorithms
based on the duality concepts is shown.
\end{abstract}



\maketitle


\section{Introduction}

Duality relations between stochastic processes have been investigated
mainly in mathematical physics,
and it has been shown that the duality concepts
are useful to investigate nonequilibrium systems \cite{Liggett_book}.
Ranging from the asymmetric simple exclusion processes (for example, see 
Ref.~\cite{Schutz1997})
to stochastic differential equations (for example, see Ref.~\cite{Shiga1986}),
the duality concepts have been widely employed.
As for the duality concepts, see a recent review paper in Ref.~\cite{Jansen2014}.

The previous works for the duality concepts are based on analytical solutions
for the dual stochastic processes;
when the original process is intractable analytically,
its tractable dual process is solved.
However, it is still unclear whether the duality concepts
can be combined with numerical methods
and work well or not. 
Recent developments for duality relations for stochastic processes
enable us to derive dual birth-death processes
from various stochastic differential equations
\cite{Giardina2009,Ohkubo2010,Ohkubo2013b,Carinci2015},
and the derived dual birth-death processes are sometimes analytically intractable.
Is it possible to utilize such analytically intractable
birth-death processes in useful ways?
In addition, is there a merit to use the duality concept?

The main aim of the present paper is to seek the possibility
for the combination of the duality concept and numerical methods.
As a demonstration, we here consider a filtering problem;
there are unobserved hidden states in stochastic differential equations,
and our task is to estimate the hidden states
only from partially observed data.
A new estimation method based on the duality concepts is proposed,
which is based on the ensemble Kalman filter (EnKF)
\cite{Evensen1994}.
As a result, we see that the method based on the duality concepts
can work at least for a simple nonlinear system.
In addition, there is a possibility that the duality concepts
enable us to make faster computational algorithms from a novel viewpoint;
we should perform numerical simulations for the dual birth-death processes
\textit{in advance},
and the numerical results can be used for the time-evolution
for the original stochastic differential equations with arbitrary initial conditions.
Hence, there is no need to perform Monte Carlo simulations
for the original stochastic differential equations
at each measurement time step;
we can \textit{reuse} the numerical results for the dual birth-death processes
repeatedly.

The present paper is constructed as follows.
In Sec.~II, the model used in the present paper is explained.
Section III is a brief review of the EnKF.
The main proposal in the present paper is given in Sec.~IV;
the derivation of the dual birth-death process
and the usage of the duality relation are explained.
In Sec.~V, results of a demonstration of the new algorithm and 
comparisons with the EnKF are given.
Section VI is for concluding remarks.

\section{Model}
\label{sec_2}

\subsection{Time-evolution of the state variables}

In the present paper, the following Van der Pol-type model,
which was used for a test of the filtering problem in
Ref.~\cite{Lakshmivarahan2009}, is considered:
\begin{align}
\begin{cases}
\frac{d}{dt} x_1(t) = x_2(t) + w_1(t),\\
\frac{d}{dt} x_2(t) = \epsilon (1-x_1(t)^2) x_2(t) - x_1(t) + w_2(t),
\end{cases}
\label{eq_model}
\end{align}
where $w_i(t) \in \mathbb{R}$ is a zero-mean white Gaussian noise with 
a covariance matrix $Q \in \mathbb{R}^{2 \times 2}$.
Different from the original Van der Pol model,
the model Eq.~\eqref{eq_model} contains the noise terms.
Here, we assume that the noises in Eq.~\eqref{eq_model} are not correlated 
with each other,
and then the covariance matrix is a diagonal matrix; 
$Q = \mathrm{diag}[Q_{11}, Q_{22}]$.
In the following,
the vector $\bm{x}(t) = [x_1(t) \,\, x_2(t)]^\mathrm{T}$ is sometimes used 
for notational brevity.

\subsection{Measurements}

The time-evolution of the state variable $\bm{x}$ obeys Eq.~\eqref{eq_model}.
Here, suppose that only one of the state variable $\bm{x}$ can be observed,
and that the measurement is performed with certain time intervals.
That is, although the time-evolution of the model Eq.~\eqref{eq_model} is continuous,
the measurement results are obtained only for the discrete times
$\{\tau_1, \tau_2, \cdots\}$.
For simplicity, in the present paper, 
we basically consider that the time interval of the measurements,
$\Delta \tau^\mathrm{obs}_k \equiv \tau_k - \tau_{k-1}$, is fixed,
i.e., $\Delta \tau^\mathrm{obs}_k = \Delta \tau^\mathrm{obs} $ for all $k$.
Note that it is easy to extend the methods discussed in the present paper
to variable time interval cases.

In summary, the following measurement procedure at time $\tau_k$ is employed:
\begin{align}
y(\tau_k) = H \bm{x}(\tau_k) + v(\tau_k),
\label{eq_measurement}
\end{align}
where $H = [0 \,\, 1]$ and $v(\tau_k)$ is a zero-mean white Gaussian noise
with variance $R$.
Hence, only some parts of the state variable $x_2$ are observed
with the addition of the measurement noise.

\subsection{Data used in the present paper}

\begin{figure}
\includegraphics[width=90mm,keepaspectratio]{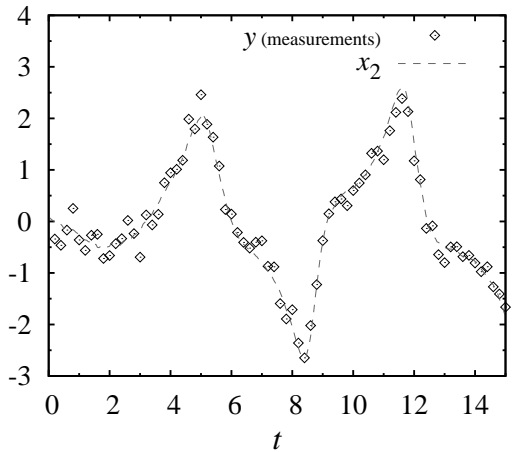}
\caption{
Measurement data.
Although a trajectory of $\bm{x}(t)$ is simulated
using Eq.~\eqref{eq_model},
the measurement is only performed for $x_2$ with the discrete time interval
$\Delta \tau^\mathrm{obs} = 0.2$.
In addition, each measurement includes measurement noise (the variance $R = 0.04$).
We also depict the original trajectory for $x_2$
as the dashed line.
}
\label{fig_1}
\end{figure}

The discrete version of the model Eq.~\eqref{eq_model} has been 
used in Ref.~\cite{Lakshmivarahan2009},
and hence we here employ the following parameters, 
which are similar to the previous work:
$\epsilon = 1.0$, $Q_{11} = 0.0262$, $Q_{22} = 0.008$, and $R = 0.04$.
(Compared with the work in Ref.~\cite{Lakshmivarahan2009},
we use a little larger measurement noise.)
Using these parameters, the data for the estimation problem
is created as follows.

Firstly, the time-evolution in Eq.~\eqref{eq_model} is simulated
using the first-order Euler-Maruyama scheme \cite{Kloeden_book,Gardinar_book};
the time interval for the simulation is $10^{-4}$,
and the initial conditions are $x_1(0) = 0.2$ and $x_2(0) = 0.1$.
Secondly, after the simulation of the state variable $\bm{x}(t)$,
the measurement procedure is performed;
only the state variable $x_2$ is extracted,
and the measurement noises are added.
The time interval for the measurements is $\Delta \tau^\mathrm{obs} = 0.2$.
Finally, we obtain the measurement data depicted in Fig.~\ref{fig_1}.

The aim of the problem here is to estimate
the state variable $\bm{x}(t)$ (i.e., not only $x_2(t)$, but also $x_1(t)$)
from the partially measured data in Fig.~\ref{fig_1}.
Although it is necessary to estimate
the covariance matrix $Q$ for the noise in the model
and the variance $R$ of the measurement noise in practice,
in the present paper,
we assume that these parameters are previously known for simplicity.

\section{Brief review of EnKF}

The most famous method for adaptive estimation of the hidden states
is the Kalman filter (as for this topic, for example, see 
Refs.~\cite{Haykin_book,Candy_book}.)
The original Kalman filter was formulated for linear systems under Gaussian noise,
and its nonlinear extensions have been studied well.
The extensions include the extended Kalman filter,
the particle filter, and the EnKF;
for the details of these filters,
for example, see Ref.~\cite{Candy_book}.

The aim of the present paper is to demonstrate
the applicability of the duality concepts.
Hence, a new numerical method for the filtering problem,
which is based on the duality concepts,
will be proposed and discussed later.
As a method for comparison, we here employ the EnKF
\cite{Evensen1994,Gillijns2006,Evensen2009}
because the new method, which will be introduced in Sec.IV,
is based on the EnKF.
Hence, we here briefly explain the EnKF.

The EnKF uses the Monte Carlo simulations;
using an ensemble of many particles,
statistical quantities such as means and covariances are evaluated,
and these quantities are employed to calculate an important quantity, so-called Kalman gain.
Note that we here use the problem settings in Sec.~II,
and hence the measurements are performed only for the discrete times
$\{\tau_1, \tau_2, \cdots \}$.

\vspace{2mm}
\noindent
\textbf{Algorithm for the EnKF}
\begin{enumerate}
\item[1.] Initialization:\\
Make the initial ensemble.
Here, we choose $n$ samples from a Gaussian distribution
with mean $\overline{\bm{x}}(0)$ and covariance matrix $P(0)$,
where $\overline{\bm{x}}(0)$ and $P(0)$ are chosen arbitrarily.
We denote each sample at time $\tau_0=0$ as
$\bm{x}_i(\tau_0)$ for $i = 1, 2, \cdots, n$.
\item[2.] Forecast step ($\bm{x}_i(\tau_{k-1}) \to \bm{x}^\mathrm{f}_i(\tau_k)$):\\
Using the time-evolution of Eq.~\eqref{eq_model},
simulate the path of the state variable for each sample
starting from $\bm{x}_i(\tau_{k-1})$.
For the simulation, for example, the first-order Euler-Maruyama scheme is available.
The simulated path for sample $i$ is denoted as
$\bm{x}^\mathrm{f}_i(t)$ ($\tau_{k-1} < t \leq \tau_k$).
\item[3.] Assimilation step ($\bm{x}^\mathrm{f}_i(\tau_k) \to \bm{x}_i(\tau_k)$):
\begin{enumerate}
\item Make realizations of random variables $\{v_i(\tau_k)\}_{i=1}^n$ as the measurement noises.
Each realization is obtained from the zero-mean white Gaussian noise with
the variance $R$.
\item Calculate following quantities:\\
(mean evaluated from the ensemble)
\begin{align}
\overline{\bm{x}}^\mathrm{f}(\tau_k) = \frac{1}{n} \sum_{i=1}^n \bm{x}^\mathrm{f}_i(\tau_k),
\end{align}
(error matrix $E^\mathrm{f}(\tau_k) \in \mathbb{R}^{2\times n}$)
\begin{align}
&E^\mathrm{f}(\tau_k) \nonumber \\
&= [\bm{x}^\mathrm{f}_1(\tau_k) - \overline{\bm{x}}^\mathrm{f}(\tau_k) \quad \cdots \quad 
\bm{x}^\mathrm{f}_n(\tau_k) - \overline{\bm{x}}^\mathrm{f}(\tau_k)],
\end{align}
(unbiased covariance evaluated from the ensemble)
\begin{align}
\widehat{P}^\mathrm{f}(\tau_k) = \frac{1}{n-1} E^\mathrm{f}(\tau_k) (E^\mathrm{f}(\tau_k))^\mathrm{T},
\end{align}
(mean of the measurement noises)
\begin{align}
\overline{v}(\tau_k) = \frac{1}{n} \sum_{i=1}^n v_i(\tau_k),
\end{align}
(unbiased variance of the measurement noise)
\begin{align}
\widehat{R}(\tau_k) = \frac{1}{n-1}
\sum_{i=1}^n \left(v_i(\tau_k) - \overline{v}(\tau_k) \right)^2,
\end{align}
(Kalman gain)
\begin{align}
\widehat{K}(\tau_k) = \widehat{P}^\mathrm{f}(\tau_k) H^\mathrm{T} 
\left( H \widehat{P}^\mathrm{f}(\tau_k) H^\mathrm{T} + \widehat{R}(\tau_k) \right)^{-1}.
\end{align}
\item[4.] Modify the forecasted state variables $\{\bm{x}^\mathrm{f}_i(\tau_k)\}_{i=1}^{n}$
using the measurement at time $\tau_k$, i.e., $y(\tau_k)$, as follows:
\begin{align}
\bm{x}_i(\tau_k) 
= \bm{x}^\mathrm{f}_i(\tau_k) + \widehat{K}(\tau_k) 
\left(
y(\tau_k) + v_i(\tau_k) - H \bm{x}^\mathrm{f}_i(\tau_k)
\right).
\end{align}
\end{enumerate}
\end{enumerate}
Steps 2 and 3 in the above algorithm are performed
for each measurement time step.

In the EnKF,
the time evolution in the forecast step 
is performed as the nonlinear systems,
which gives the non-Gaussian distribution for $\{\bm{x}_i (t)\}_{i=1}^n$
even if we start from a Gaussian distribution.
After the time evolution,
at each assimilation step,
the conventional Kalman filter is employed,
which means a filter at least up to the second moment.
The final filtered value of the state variable $\bm{x}(t)$
is obtained as the mean of the ensemble $\{\bm{x}_i(t)\}_{i=1}^n$.
In addition, the estimated error could be obtained
from the (co)variance of the ensemble $\{\bm{x}_i(t)\}_{i=1}^n$.

One of the problem in the EnKF is as follows:
in order to obtain more accurate estimation results,
we need large ensemble size $n$.
That is, a small ensemble size gives
an inaccurate Kalman gain, and hence
the estimation results would not be accurate.
In general,
it is easy to imagine that the large ensemble size needs high computational costs.
Although it has been clarified that 
a small ensemble size is often enough for the EnKF 
in practical cases \cite{Lakshmivarahan2009},
it would be preferable if we could avoid the numerical simulations
of the time evolution of the ensemble
at each measurement time step.

\section{Algorithm based on duality relations}

\subsection{Basic concept}

Here, a simple explanation for the basic concept of the duality relation is given;
we consider here the duality relation between
a stochastic differential equation and a birth-death process.
For simplicity, stochastic processes with only one random variable
are considered in this subsection.

Suppose that $(x_t)_{t \ge 0} \in \mathbb{R}$ is a sample trajectory 
of the stochastic differential equation,
and $p(x,t)$ is the probability density at time $t$.
It has been known that
some stochastic differential equations
have the corresponding dual birth-death processes.
Denote a dual birth-death process as $(n_t)_{t \ge 0} \in \mathbb{N}$,
whose probability distribution at time $t$ is $P(n,t)$,
and then the following equality is satisfied 
\cite{Liggett_book,Giardina2009,Ohkubo2010,Ohkubo2013b}:
\begin{align}
\mathbb{E}_{x}\left[ x_t^{n_0} \right]
= \mathbb{E}_{n}\left[ x_0^{n_t} \right],
\label{eq_simple_duality}
\end{align}
where $\mathbb{E}_x$ and $\mathbb{E}_n$
are the expectations in the stochastic differential equation $(x_t)_{t \ge 0}$ starting from $x_0$
and in the birth-death process $(n_t)_{t \ge 0}$ starting from $n_0$, respectively.
More explicitly, we can rewrite Eq.~\eqref{eq_simple_duality} as
\begin{align}
\int_{- \infty}^{\infty} p(x,t) x^{n_0} dx = \sum_{n=0}^\infty P(n,t) x_0^n,
\end{align}
where $p(x,0) = \delta(x - x_0)$
and $P(n,0) = \delta_{n,n_0}$.

Equation~\eqref{eq_simple_duality} shows that
the information about the stochastic differential equation
can be obtained from the solution of the birth-death process.
That is, when we obtain the probability distribution of the birth-death process, $P(n,t)$,
with the initial condition $n_0 = 1$,
it is possible to evaluate the first order moment, i.e., the mean value of $x_t$,
of the stochastic differential equation,
without solving the stochastic differential equation.

There are several advantages of the usage of the duality relations.
It is sometimes easier to treat the birth-death process,
compared with the stochastic processes.
For some specific cases, the analytical solution of the birth-death process
has been obtained.
In addition, there are numerical algorithms to simulate the birth-death process efficiently.
As for numerical evaluations of the stochastic differential equations,
we need some approximation; for example,
the time-discretization is needed in the Euler-Maruyama scheme.
On the other hand, for example, the Gillespie algorithm 
for the birth-death process does not need the time-discretization \cite{Gillespie1977}.
In this sense, the birth-death process would be more tractable 
than the stochastic differential equations.

Furthermore, 
only `\textit{a}' solution of the birth-death process
can be used to obtain the information 
about the stochastic differential equations
with `\textit{arbitrary}' initial conditions.
That is, if $n_0 = 1$,
$\mathbb{E}_{x} [ x_t ] = \mathbb{E}_{n} [ x_0^{n_t}] = \sum_{n=0}^\infty P(n,t) x_0^n$,
and $P(n,t)$ is independent of the value of $x_0$.
Hence, using only a solution of the birth-death process, $P(n,t)$,
it is possible to estimate the average of $x_t$, $\mathbb{E}_{x} [ x_t ]$,
for the stochastic differential equations
for `arbitrary' initial conditions $x_0$.
Of course, if we want to know the first and second moments of $x_t$ in Eq.~\eqref{eq_simple_duality},
two solutions of the birth-death process $n_t$
with the different initial conditions, $n_0 = 1$ and $n_0 = 2$,
are needed.
However, once we have these two solutions of the birth-death process,
the first and second moments, $\mathbb{E}_x[x_t]$ and $\mathbb{E}_x[x_t^2]$, 
with arbitrary initial conditions can be evaluated
by using the duality relation in Eq.~\eqref{eq_simple_duality}.
In contrast, when these moments are evaluated from the direct simulation
of the stochastic differential equation,
we need many sample trajectories with an initial condition $x_0$;
if the initial condition $x_0$ is changed,
we must perform many other numerical simulations.

The basic idea for a new algorithm is as follows;
the Monte Carlo simulation in the forecast step
in the EnKF is replaced with
the simple numerical evaluation based on the duality relation.

The remaining problem is as follows:
How should we derive the dual birth-death process
from a given stochastic differential equation?
In the successive subsections,
we will show the method to obtain the dual birth-death process,
employing a mathematical formalism
called the Doi-Peliti formalism \cite{Doi1976a,Doi1976b,Peliti1985,Tauber2005}.

\subsection{Doi-Peliti formalism}

We here briefly review the Doi-Peliti formalism,
which is useful to obtain the duality relations.
The Doi-Peliti formalism is the method similar to the second quantization method
in quantum mechanics.
Up to now, the Doi-Peliti formalism has been used in various contexts,
mainly in order to investigate discrete systems such as chemical reactions,
and it has been shown that the algebraic probability theory \cite{Hora_book}
gives the mathematical basis of the Doi-Peliti formalism
\cite{Ohkubo2013a}.

In the Doi-Peliti formalism,
creation operator $a^\dagger$ and annihilation operator $a$ are introduced,
which satisfy the following commutation relation:
\begin{align}
[a, a^\dagger] \equiv a a^\dagger - a^\dagger a = 1, \quad
[a, a] = [a^\dagger, a^\dagger] = 0.
\label{eq_commutation_relation}
\end{align}
These operators act on a vector in the Fock space, $|n\rangle$, as follows:
\begin{align}
a^\dagger | n \rangle = | n+1 \rangle, \quad
a | n \rangle = n | n-1 \rangle,
\end{align}
and the vacuum state $| 0 \rangle$ is characterized by $a | 0 \rangle = 0$.
Additionally, vectors $\{\langle m |\}_{m = 0}^{\infty}$ satisfy
the following orthogonal relation to the vectors $\{|n\rangle\}_{n=0}^\infty$:
\begin{align}
\langle m | n \rangle = \delta_{m,n} n!.
\end{align}
Note that $a^\dagger a$ corresponds to the number operator
and
\begin{align}
a^\dagger a | n \rangle = n | n \rangle.
\label{eq_number_operator}
\end{align}

It has been shown that the Doi-Peliti formalism
is deeply related to the conventional generating function method \cite{Droz1994},
and the following correspondences can be useful
to understand the usage of the Doi-Peliti formalism in the duality problem:
\begin{align}
| n \rangle \leftrightarrow x^n,
\quad
a^\dagger \leftrightarrow x, \quad a \leftrightarrow \frac{\partial}{\partial x}.
\label{eq_Doi_Peliti_correspondence}
\end{align}
That is, the differential operator is connected to the annihilation operator,
which is used in partial differential equations derived from stochastic differential equations.
In addition, the annihilation operator acts on the \textit{discrete} states $| n \rangle$,
which is available to construct the birth-death process,
as shown later.
The important point here is that the Doi-Peliti formalism can bridge continuous states
with discrete states,
which corresponds to the connection
between a stochastic differential equation and a birth-death process.

\subsection{Derivation of the dual birth-death process}

The derivation of the simple duality relation, such as Eq.~\eqref{eq_simple_duality}, 
has been discussed in Ref.~\cite{Giardina2009},
and the derivation based on the Doi-Peliti formalism has also been proposed  \cite{Ohkubo2010}.
However, in order to treat the estimation problem,
only the simple duality relation in Eq.~\eqref{eq_simple_duality} is not enough;
the simple duality relation in Eq.~\eqref{eq_simple_duality}
can deal with only a very restricted class of stochastic differential equations.
Recently, extended duality relations have been proposed \cite{Ohkubo2013b},
which is necessary to construct the new algorithm based on the duality relation.
Here, we only show, as an example, the derivation of a dual birth-death process
from the stochastic differential equations in Eq.~\eqref{eq_model}.
For the mathematical details, see the original paper \cite{Ohkubo2013b}.

First of all, it is needed to construct
the corresponding Fokker-Planck equation of the stochastic differential equations in Eq.~\eqref{eq_model}.
(For the derivation of the Fokker-Planck equation
from the stochastic differential equations,
see, for example, Ref.~\cite{Gardinar_book}.)
The corresponding Fokker-Planck equation is as follows:
\begin{align}
&\frac{\partial}{\partial t} p(x_1, x_2, t) \nonumber \\
&= \left[
-\frac{\partial}{\partial x_1} x_2 - \frac{\partial}{\partial x_2} \left[ \epsilon(1-x_1^2) x_2 - x_1 \right] 
\right] p(x_1, x_2, t) \nonumber \\
& \quad + \frac{1}{2} \left[ 
Q_{11} \frac{\partial^2}{\partial x_1^2}
+ Q_{22} \frac{\partial^2}{\partial x_2^2}
\right] p(x_1, x_2, t),
\label{eq_FP_original}
\end{align}
where $p(x_1, x_2, t)$ is the probability density at time $t$
for the stochastic differential equations.
For notational convenience,
we introduce the following linear operator $L^{*}$:
\begin{align}
&L^{*} \left( x_1,\frac{\partial}{\partial x_1}, x_2, \frac{\partial}{\partial x_2} \right) \nonumber \\
&=
\left[
-\frac{\partial}{\partial x_1} x_2 - \frac{\partial}{\partial x_2} \left[ \epsilon(1-x_1^2) x_2 -x_1 \right] 
\right] \nonumber \\
&\quad + \frac{1}{2} \left[ 
Q_{11} \frac{\partial^2}{\partial x_1^2}
+ Q_{22} \frac{\partial^2}{\partial x_2^2}
\right],
\label{eq_FP_operator}
\end{align}
and hence the Fokker-Planck equation ~\eqref{eq_FP_original}
is rewritten as
\begin{align}
\frac{\partial}{\partial t} p(x_1, x_2, t)
= 
L^{*} \left( x_1,\frac{\partial}{\partial x_1}, x_2, \frac{\partial}{\partial x_2} \right)
p(x_1, x_2, t).
\label{eq_FP}
\end{align}
The adjoint operator of Eq.~\eqref{eq_FP_operator}, $L$, is as follows:
\begin{align}
&L  \left( x_1,\frac{\partial}{\partial x_1}, x_2, \frac{\partial}{\partial x_2} \right) \nonumber \\
&= 
x_2 \frac{\partial}{\partial x_1} 
- \epsilon(1-x_1^2) x_2  \frac{\partial}{\partial x_2}  
- x_1 \frac{\partial}{\partial x_2} \nonumber \\
&\quad + \frac{1}{2} \left[ 
Q_{11} \frac{\partial^2}{\partial x_1^2}
+ Q_{22} \frac{\partial^2}{\partial x_2^2}
\right],
\end{align}
and using the correspondence
between operators in the Doi-Peliti formalism
and the differential operators in Eq.~\eqref{eq_Doi_Peliti_correspondence},
we have
\begin{align}
&L \left( a^\dagger_1, a_1, a^\dagger_2, a_2 \right) \nonumber \\
&= 
a^\dagger_2 a_1  +  \epsilon (1- a^\dagger_1 a^\dagger_1) a^\dagger_2 a_2  - a^\dagger_1 a_2  \nonumber \\
&\quad + \frac{1}{2} \left[ 
Q_{11} a_1 a_1  + Q_{22} a_2 a_2
\right],
\end{align}
where the following correspondences are used:
\begin{align}
a_1^\dagger \leftrightarrow x_1, \quad a_1 \leftrightarrow \frac{\partial}{\partial x_1}, 
\quad
a_2^\dagger \leftrightarrow x_2, \quad a_2 \leftrightarrow \frac{\partial}{\partial x_2}.
\label{eq_correspondence_1}
\end{align}
In addition, as discussed later,
it is convenient to introduce the time scaling $t = r_\mathrm{ts} \tilde{t}$;
due to the time scaling,
the original Fokker-Planck equation ~\eqref{eq_FP} is rewritten as
\begin{align}
&\frac{\partial}{\partial \tilde{t}} p(x_1, x_2, r_\mathrm{ts} \tilde{t})  \nonumber \\
&= 
r_\mathrm{ts} L^{*} \left( x_1,\frac{\partial}{\partial x_1}, x_2, \frac{\partial}{\partial x_2} \right)
p(x_1, x_2, r_\mathrm{ts} \tilde{t}).
\end{align}
We here focus on the fact that 
continuous variables, $x_1$ and $x_2$, are replaced with creation operators
in Eq.~\eqref{eq_correspondence_1}.
Hence, if we reinterpret a constant as a creation operator,
the following replacement is available:
\begin{align}
r_\mathrm{ts} \leftrightarrow a^\dagger_0,
\label{eq_correspondence_2}
\end{align}
and therefore the following linear operator is obtained:
\begin{align}
&L \left( a^\dagger_0, a^\dagger_1, a_1, a^\dagger_2, a_2 \right)  \nonumber \\
&\equiv a^\dagger_0 L \left( a^\dagger_1, a_1, a^\dagger_2, a_2 \right) \nonumber \\
&= 
a^\dagger_0 a^\dagger_2 a_1  
+ \epsilon a^\dagger_0 a^\dagger_2 a_2 
- \epsilon a^\dagger_0 (a^\dagger_1)^2 a^\dagger_2 a_2  - a^\dagger_0 a^\dagger_1 a_2 \nonumber \\
&\quad + \frac{1}{2} \left[ 
Q_{11} a^\dagger_0 a_1 a_1 
+ Q_{22} a^\dagger_0 a_2 a_2
\right].
\label{eq_L}
\end{align}
Note that all creation operators in each term in Eq.~\eqref{eq_L} are placed 
on the left side of the annihilation operators;
if not, we must replace the term using the commutation relation in Eq.~\eqref{eq_commutation_relation}.

Since the linear operator in Eq.~\eqref{eq_L}
acts on the Fock space, i.e., the discrete state space $\{|n \rangle\}_{n=0}^\infty$,
one may expect that the linear operator simply gives
the time-evolution for a birth-death process.
That is, defining the state vector 
$| \psi (\tilde{t})\rangle$ as
\begin{align*}
| \psi (\tilde{t})\rangle \equiv 
\sum_{n_0 = 0}^\infty \sum_{n_1=0}^\infty \sum_{n_2 = 0}^\infty 
P(n_0, n_1, n_2, \tilde{t}) | n_0, n_1, n_2 \rangle,
\end{align*}
where $P(n_0, n_1, n_2, \tilde{t})$ is a probability distribution of a birth-death process,
and considering the time-evolution equation
\begin{align*}
\frac{\partial}{\partial \tilde{t}} | \psi(\tilde{t}) \rangle
= L(a_0^\dagger, a_1^\dagger, a_1, a_2^\dagger, a_2) | \psi(\tilde{t}) \rangle,
\end{align*}
we may have a time-evolution equation for $P(n_0, n_1, n_2, \tilde{t})$
by comparing the coefficient of a state vector $| n_0, n_1, n_2 \rangle$
on the right and left hand sides.
However, as discussed in the previous work \cite{Ohkubo2013b},
it is impossible to simply interpret the linear operator in Eq.~\eqref{eq_L} 
as a time-evolution operator for a birth-death process;
the linear operator does not satisfy the probability conservation law,
and, in addition,
some terms seem to correspond to `negative' transition rates.

In order to construct an adequate birth-death process,
we need an additional operator $b$, which satisfies
the following relations:
\begin{align}
b | + \rangle = | - \rangle, \quad b | - \rangle = | + \rangle,
\end{align}
where $| + \rangle$ and $| - \rangle$ are orthonormal state vectors satisfying
$\langle + | + \rangle = \langle - | - \rangle = 1$
and $\langle + | - \rangle = \langle - | + \rangle = 0$.
The following procedure is needed to construct
an adequate time-evolution operator for a birth-death process 
(although the following procedure might seem complicated, 
a concrete example will be given soon):
\begin{enumerate}
\item[] 
For each term in the linear operator $L$,
apply the following procedures.
\item[1.] If the coefficient of the term has a negative sign,
replace the negative sign `$-$' with `$+ b$' using the operator $b$.
\item[2.] Act the operators on $| \bm{n} \rangle$, and evaluate
the coefficient including the effects of the number operators
$a^\dagger_i a_i$.
\item[3.] Subtract a term in order to guarantee the probability conservation law;
the term gives the same coefficient with Step 2,
and additionally, the term consists of 
the same number of creation and annihilation operators
for the same sub-index.
\item[4.] In order to compensate the subtracted term,
add the same term with Step 3.
\end{enumerate}
For example, the first term in Eq.~\eqref{eq_L} is $a^\dagger_0 a^\dagger_2 a_1$,
and then we do not need Step 1.
Since
\begin{align}
a^\dagger_0 a^\dagger_2 a_1 | n_0, n_1, n_2 \rangle
= n_1 | n_0+1, n_1-1, n_2+1 \rangle,
\label{eq_first_term}
\end{align}
the coefficient is $n_1$,
and therefore the term $a^\dagger_1 a_1$ must be subtracted in Step 2.
On the other hand, the third term in Eq.~\eqref{eq_L}, $- \epsilon a^\dagger_0 (a^\dagger_1)^2 a^\dagger_2 a_2$,
has the negative sign,
and hence Step 1 gives 
$+ b \epsilon a^\dagger_0 (a^\dagger_1)^2 a^\dagger_2 a_2$.
The subtracted term in Step 3 is $b \epsilon a^\dagger_2 a_2$
because
\begin{align}
a^\dagger_0 (a^\dagger_1)^2 a^\dagger_2 a_2 | n_0, n_1, n_2 \rangle
= n_2 | n_0+1, n_1+2, n_2 \rangle,
\label{eq_third_term}
\end{align}
which gives the coefficient $n_2$.

Using the above procedure,
we obtain the following linear operator,
instead of the linear operator $L$ in Eq.~\eqref{eq_L}:
\begin{align}
&L \left( a^\dagger_0, a_0, a^\dagger_1, a_1, a^\dagger_2, a_2, b \right) \nonumber \\
&= 
L' \left( a^\dagger_0, a_0, a^\dagger_1, a_1, a^\dagger_2, a_2, b \right) 
+ V \left( a^\dagger_1 a_1, a^\dagger_2 a_2 \right),
\label{eq_new_L}
\end{align}
where
\begin{align}
&L' \left( a^\dagger_0, a_0, a^\dagger_1, a_1, a^\dagger_2, a_2, b \right) \nonumber \\
&= 
(a^\dagger_0 a^\dagger_2 a_1  - a^\dagger_1 a_1)
+ (\epsilon a^\dagger_0 a^\dagger_2 a_2  - \epsilon a^\dagger_2 a_2) \nonumber \\
&\quad + (\epsilon b a^\dagger_0 (a^\dagger_1)^2 a^\dagger_2 a_2
- \epsilon a^\dagger_2 a_2)
+ (b a^\dagger_0 a^\dagger_1 a_2 - a^\dagger_2 a_2) \nonumber \\
&\quad + \left( \frac{1}{2} Q_{11} a^\dagger_0 a_1 a_1 - \frac{1}{2} Q_{11} a^\dagger_1 a^\dagger_1 a_1 a_1 \right) \nonumber \\
&\quad + \left( \frac{1}{2} Q_{22} a^\dagger_0 a_2 a_2 -  \frac{1}{2} Q_{22} a^\dagger_2 a^\dagger_2 a_2 a_2 \right)
\label{eq_Lp}
\end{align}
and
\begin{align}
&V \left( a^\dagger_1 a_1, a^\dagger_2 a_2 \right)  \nonumber \\
&= a^\dagger_1 a_1 + 2 \epsilon a^\dagger_2 a_2
+ a^\dagger_2 a_2 
+ \frac{1}{2} Q_{11} \left( (a^\dagger_1 a_1) (a^\dagger_1 a_1) - a^\dagger_1 a_1 \right) \nonumber \\
&\quad + \frac{1}{2} Q_{22} \left( (a^\dagger_2 a_2) (a^\dagger_2 a_2) - a^\dagger_2 a_2 \right).
\label{eq_V}
\end{align}
Note that we used $a^\dagger a^\dagger a a = (a^\dagger a)(a^\dagger a) - a a$,
which stems from $[a,a^\dagger] = a a^\dagger - a^\dagger a = 1$.

As shown in the previous work \cite{Ohkubo2013b},
it is possible to interpret the linear operator $L'$ in Eq.~\eqref{eq_Lp}
as a time-evolution operator for a birth-death process.
Because of the operator $b$,
we must consider an additional state variable, which takes only two states ($+$ or $-$),
in addition to the state variables $n_0, n_1$ and $n_2$.
See the first term in Eq.~\eqref{eq_Lp};
the action of the first term on $| \bm{n} \rangle$ is Eq.~\eqref{eq_first_term},
and hence we can interpret this term
as an elementary birth-death process
$X_1 \to X_2 + X_0$ with rate $n_1$,
where $n_0, n_1$ and $n_2$ is the number of particles $X_0, X_1$, and $X_2$, respectively.
On the other hand, the third term in Eq.~\eqref{eq_Lp}
gives $X_2 \to 2 X_1 + X_2 + X_0$
and the state change with $+ \to -$ or $- \to +$
with rate $\epsilon n_2$;
the state ($+$ or $-$) is changed due to an event corresponding to the third term.
Repeating the similar discussions,
finally the following birth-death process is obtained:
\begin{align}
\begin{cases}
\mathrm{(i)} &X_1 \xrightarrow{n_1} X_2 + X_0, \\
\mathrm{(ii)} &X_2 \xrightarrow{\epsilon n_2} X_2 + X_0, \\
\mathrm{(iii)} &X_2 \xrightarrow{\epsilon n_2} 2 X_1 + X_2 + X_0 \quad \textrm{with S.C.},\\
\mathrm{(iv)} &X_2 \xrightarrow{n_2} X_1 + X_0 \quad \textrm{with S.C.},\\
\mathrm{(v)} &2 X_1 \xrightarrow{n_1(n_1-1)} X_0, \\
\mathrm{(vi)} &2 X_2 \xrightarrow{n_2(n_2-1)} X_0,
\end{cases}
\label{eq_birth_death}
\end{align}
where `S.C.' means the state change $+ \to -$ or $- \to +$.


The term $V(a^\dagger_1 a_1, a^\dagger_2 a_2)$
is called a Feynman-Kac term,
and the important point is that this term is written
only in terms of the number operators $a^\dagger_i a_i$.
Since the number operators $a^\dagger_i a_i$
does not affect the state vectors,
we can simply replace the number operators with the random variables
in the birth-death process; i.e.,
$V ( a^\dagger_1 a_1, a^\dagger_2 a_2 ) = V(n_1,n_2)$.

The intuitive understanding of the duality relation is written
as follows:
we here abbreviate a state related to
the Fokker-Plank equation at time $t$ as $\mathrm{FP}(t)$
and that to the birth-death process
as $\mathrm{BD}(t)$.
In addition, for simplicity, set $r_\mathrm{ts} = 1$ and hence $t = \tilde{t}$;
furthermore, we neglect the Feynman-Kac term here.
Then, formally, time-evolution of 
$\mathrm{BD}(t)$ is given by $\mathrm{BD}(t) = e^{L t}\mathrm{BD}(0)$.
In contrast, that of $\mathrm{FP}(t)$ 
is written as $\mathrm{FP}(t) = \mathrm{FP}(0) e^{L t}$,
when we consider the left-action of $L$.
Hence, the adjoint operator $L^{*}$ gives the actual time-evolution
of the Fokker-Planck equation.
In addition, we have formally
$\mathrm{FP}(t) \mathrm{BD}(0)= \mathrm{FP}(0) e^{Lt} \mathrm{BD}(0)= \mathrm{FP}(0)  \mathrm{BD}(t)$;
this corresponds to the duality relation.
The linear operator $L$ can be written
in terms of both the differential operators and the creation-annihilation operators,
and hence the stochastic differential equation in Eq.~\eqref{eq_model}
and the birth-death process in Eq.~\eqref{eq_birth_death} are connected naturally.
Of course, the above discussion is just an intuitive one,
and for the mathematical explanation of the duality relations,
see the previous work \cite{Ohkubo2013b}.

Here we consider a time-evolution from time $0$ to $\tau$.
Because we consider the time-scaling factor $r_\mathrm{ts}$,
the time interval, $\tau$, in the stochastic differential equations
correspond to the time interval $\tilde{\tau} = \tau / r_\mathrm{ts}$
in the dual birth-death process.
Using the duality relation,
we can obtain the following identity:
\begin{widetext}
\begin{align}
&\mathbb{E}_{\bm{x}}
\left[
x_1(\tau)^{n_1(0)} x_2(\tau)^{n_2(0)} 
\right] \nonumber \\
&= \mathbb{E}_{\bm{n},+} \Bigg[
\exp\left\{
\int_{0}^{\tilde{\tau}} V(n_1(s),n_2(s)) ds
\right\}
 r_\mathrm{ts}^{n_0(\tilde{\tau})} 
x_1(0)^{n_1(\tilde{\tau})} x_2(0)^{n_2(\tilde{\tau})} 
\Bigg]
\nonumber \\
& \,\, - \mathbb{E}_{\bm{n},-} \Bigg[
\exp\left\{
\int_{0}^{\tilde{\tau}} V(n_1(s),n_2(s)) ds
\right\}
 r_\mathrm{ts}^{n_0(\tilde{\tau})} 
x_1(0)^{n_1(\tilde{\tau})} x_2(0)^{n_2(\tilde{\tau})} 
\Bigg],
\label{eq_final_duality_delta}
\end{align}
\end{widetext}
where the abbreviations 
$\bm{x} = (x_1, x_2)$ and $\bm{n} = (n_0, n_1, n_2)$ are used;
we set $n_0(0) = 0$;
$\mathbb{E}_{\bm{n},+}$ and $\mathbb{E}_{\bm{n},-}$ are
the expectations related to the states $+$ and $-$
in the birth-death process in Eq.~\eqref{eq_birth_death}, respectively.
Using the duality relation in Eq.~\eqref{eq_final_duality_delta},
we can evaluate various information for the stochastic differential equation in Eq.~\eqref{eq_model}
from the solution of the birth-death process in Eq.~\eqref{eq_birth_death}.
In order to evaluate the probability distribution and
the corresponding Feynman-Kac terms for the birth-death process in Eq.~\eqref{eq_birth_death},
Monte Carlo simulations with the Gillespie algorithm
are available.

As noted in Sec.~IV.A,
the initial conditions for the birth-death process
correspond to the order of moments
in the stochastic differential equation.
When we use 
$n_1(0) = 2$
and $n_2(0) = 0$,
the time-evolution
of the second moment for $x_1$,
i.e., $\mathbb{E}_{\bm{x}} [x_1(\tau)^2]$,
is evaluated.

Using the above procedures,
the time evolution of the stochastic differential equation
\textit{for arbitrary initial conditions}
can be replaced with that of the dual birth-death process
\textit{for specific initial conditions}.
In addition, if changing the time-scaling variable
$r_\mathrm{ts}$,
we can evaluate information
for various time interval in the stochastic differential equations
from only a result for a single fixed time interval $\tilde{\tau}$
in the birth-death process.
For example, assume that $\tilde{\tau} = 1$;
if we want to know the information of the stochastic differential equation
at time $t = 1$,
we should set $r_\mathrm{ts} = 1$.
On the other hand,
the information at time $t = 0.9$ can be evaluated
from the same results of the birth-death process with $\tilde{\tau} = 1$ 
and setting $r_\mathrm{ts} = 0.9$.
Hence, there is no need to perform the Monte Carlo simulations
for various different time intervals;
this is also one of the advantages of the new method.

At the end of this subsection, we give one comment for the operator $b$.
In the above construction, we replaced the minus sign `$-$' with the operator `$+ b$' 
and avoided the `negative' transition problem.
Instead of that, we can use the following trick: Interpret constants as random variables.
That is, we interpret `$-1$' as a random variable $x_3$, and
the (stochastic) differential equation for $x_3$ is set as $\displaystyle d x_3/dt = 0$.
Then, employing the same discussion above, $x_3$ is replaced with $a_3^\dagger$ 
in the Doi-Peliti formulation,
and finally we obtain the following birth-death process
\begin{align}
\begin{cases}
\mathrm{(i)} &X_1 \xrightarrow{n_1} X_2 + X_0, \\
\mathrm{(ii)} &X_2 \xrightarrow{\epsilon n_2} X_2 + X_0, \\
\mathrm{(iii)} &X_2 \xrightarrow{\epsilon n_2} 2 X_1 + X_2 + X_0 + X_3,\\
\mathrm{(iv)} &X_2 \xrightarrow{n_2} X_1 + X_0 +X_3,\\
\mathrm{(v)} &2 X_1 \xrightarrow{n_1(n_1-1)} X_0, \\
\mathrm{(vi)} &2 X_2 \xrightarrow{n_2(n_2-1)} X_0,
\end{cases}
\label{eq_comment_different_birth_death}
\end{align}
and the following duality relation
\begin{align}
&\mathbb{E}_{\bm{x}}
\left[
x_1(\tau)^{n_1(0)} x_2(\tau)^{n_2(0)} x_3(\tau)^{n_3(0)} 
\right] \nonumber \\
&= \mathbb{E}_{\bm{n}} \Bigg[
\exp\left\{
\int_{0}^{\tilde{\tau}} V(n_1(s),n_2(s)) ds
\right\}
 \nonumber \\
&\qquad \quad \times
r_\mathrm{ts}^{n_0(\tilde{\tau})}
x_1(0)^{n_1(\tilde{\tau})} x_2(0)^{n_2(\tilde{\tau})} x_3(0)^{n_2(\tilde{\tau})} 
\Bigg],
\label{eq_comment_final_duality_delta}
\end{align}
where the abbreviations 
$\bm{x} = (x_1, x_2, x_3)$ and $\bm{n} = (n_0, n_1, n_2, n_3)$ are used.
Note that the Feynman-Kac term does not depend on $n_3(t)$.
Because the `random' variable $x_3(t)$ is a time-independent constant 
and $x_3(t) \equiv -1$,
we can see that the initial condition, $n_0(0)=0$ and $n_3(0)= 0$,
recovers the original duality relation in Eq.~\eqref{eq_final_duality_delta}.
This technique, in which constants are interpreted as random variables,
would be sometimes useful to consider more complicated stochastic differential equations.
However, in the example considered here, only the `negative' transition rate should be avoided,
and the expression in Eq.~\eqref{eq_final_duality_delta} is simpler in computational viewpoint;
although $n_3(t)$ takes any natural numbers, the additional states $|+\rangle$ and $|-\rangle$
takes only two states, and hence the computational memory in Monte Carlo simulations is largely reduced.
From this reason, the additional operator $b$ is introduced and used in the present paper.

\subsection{Moment evaluation}

The duality relation in Eq.~\eqref{eq_final_duality_delta}
can be used for evaluating the moments in stochastic differential equation in Eq.~\eqref{eq_model}
starting from the Dirac-delta-type initial conditions,
as discussed in Sec.~IV.A.
However, in the EnKF,
we use an ensemble of samples.
In the EnKF, only the mean and covariance matrix
of the ensemble are needed, so we can consider
that the ensemble is essentially characterized 
by a Gaussian distribution.
Hence, the Dirac-delta-type initial conditions are not enough.

We here write the expectation value, which is taken
by using a Gaussian distribution 
with mean $\bm{x}(t)$ and covariance matrix $V$,
as $\langle \cdots \rangle$.
Hence, the first term in the r.h.s. in Eq.~\eqref{eq_final_duality_delta}
should be replaced with
\begin{align*}
&
\left\langle \mathbb{E}_{\bm{n},+} \left[
e^{\int_{0}^{\tilde{\tau}} V ds}
r_\mathrm{ts}^{n_0(\tilde{\tau})} x_1(0)^{n_1(\tilde{\tau})} 
x_2(0)^{n_2(\tilde{\tau})} 
\right] \right\rangle \nonumber \\
&=
\mathbb{E}_{\bm{n},+} \left[
e^{\int_{0}^{\tilde{\tau}} V ds}
r_\mathrm{ts}^{n_0(\tilde{\tau})} \left\langle 
x_1(0)^{n_1(\tilde{\tau})} x_2(0)^{n_2(\tilde{\tau})} 
\right\rangle 
\right].
\end{align*}
We need a little additional calculations 
in order to evaluate the expectation with the Gaussian distribution;
introducing the following notations,
\begin{align}
\mu_1 = \langle x_1 \rangle,
\mu_2 = \langle x_2 \rangle,
\end{align}
the following recursion formula is known \cite{Willink2005}:
\begin{align}
\left\langle x_1^{n_1} x_2^{n_2} \right\rangle
=& \mu_1 \left\langle x_1^{n_1-1} x_2^{n_2} \right\rangle
+ V_{11} (n_1-1) \left\langle x_1^{n_1-2} x_2^{n_2} \right\rangle \nonumber \\
&+ V_{12} n_2 \left\langle x_1^{n_1-1} x_2^{n_2-1} \right\rangle 
\end{align}
or
\begin{align}
\left\langle x_1^{n_1} x_2^{n_2} \right\rangle
=& \mu_2 \left\langle x_1^{n_1} x_2^{n_2-1} \right\rangle
+ V_{21} n_1 \left\langle x_1^{n_1-1} x_2^{n_2-1} \right\rangle \nonumber \\
&+ V_{22} (n_2-1) \left\langle x_1^{n_1} x_2^{n_2-2} \right\rangle.
\end{align}
Using the above recursion formula,
once we have the probability distribution $P(n_0, n_1, n_2, t)$
for the birth-death process in Eq.~\eqref{eq_birth_death}
with adequate initial values,
it is possible to evaluate
various moments for the stochastic differential equation
in Eq.~\eqref{eq_model}
with a Gaussian initial distribution,
using the following duality relation:
\begin{widetext}
\begin{align}
&\mathbb{E}_{\bm{x}, \textrm{Gaussian initial}}
\left[
x_1(\tau)^{n_1(0)} x_2(\tau)^{n_2(0)} 
\right] \nonumber \\
&= \mathbb{E}_{\bm{n},+} \Bigg[
\exp\left\{
\int_{0}^{\tilde{\tau}} V(n_1(s),n_2(s)) ds
\right\}  
r_\mathrm{ts}^{n_0(\tilde{\tau})} 
\left\langle 
x_1(0)^{n_1(\tilde{\tau})} x_2(0)^{n_2(\tilde{\tau})} 
\right\rangle
\Bigg]
\nonumber \\
& \,\, - \mathbb{E}_{\bm{n},-} \Bigg[
\exp\left\{
\int_{0}^{\tilde{\tau}} V(n_1(s),n_2(s)) ds
\right\}
 r_\mathrm{ts}^{n_0(\tilde{\tau})} 
\left\langle
x_1(0)^{n_1(\tilde{\tau})} x_2(0)^{n_2(\tilde{\tau})} 
\right\rangle
\Bigg].
\label{eq_final_duality}
\end{align}
\end{widetext}

\subsection{Algorithm}

We here consider general cases with variable time intervals,
and denote the maximum of the time intervals,
$\tau_{k-1} - \tau_{k}$, as $\Delta \tau$. 
The new Kalman filter based on the duality relation,
called the DuKF here,
is as follows:

\vspace{2mm}
\noindent
\textbf{Algorithm for DuKF}
\begin{enumerate}
\item[1.] Preparation for the duality relations:\\
Simulate the dual birth-death process in Eq.~\eqref{eq_birth_death}.
We need simulations with five different initial conditions;
\begin{itemize}
\item[(c1)] $n_0 = 0, n_1 = 1, n_2 = 0$,
\item[(c2)] $n_0 = 0, n_1 = 0, n_2 = 1$,
\item[(c3)] $n_0 = 0, n_1 = 2, n_2 = 0$,
\item[(c4)] $n_0 = 0, n_1 = 0, n_2 = 2$,
\item[(c5)] $n_0 = 0, n_1 = 1, n_2 = 1$,
\end{itemize}
which are necessary to evaluate
$\mathbb{E}_{\bm{x}} [x_1(t)]$,
$\mathbb{E}_{\bm{x}} [x_2(t)]$,
$\mathbb{E}_{\bm{x}} [x_1(t)^2]$,
$\mathbb{E}_{\bm{x}} [x_2(t)^2]$,
$\mathbb{E}_{\bm{x}} [x_1(t) x_2(t)]$,
respectively.
For all cases, the additional state variable
is set to `$+$' initially.
The Monte Carlo simulations are performed
from $t = 0$ to $t = \Delta \tau$,
and the integral of the Feynman-Kac term $V(n_1,n_2)$
and the final probability distribution
$P(n_0,n_1,n_2,\Delta \tau)$ are evaluated numerically.
\item[2.] Initialization:\\
Set an initial Gaussian distribution
with mean $\overline{\bm{x}}^\mathrm{a}(0)$ and covariance matrix $P^\mathrm{a}(0)$.
\item[3.] Forecast step at $\tau_{k-1}$:\\
Using the duality relation in Eq.~\eqref{eq_final_duality},
evaluate various moments,
which are necessary to characterize a Gaussian distribution.
The ensemble average in Eq.~\eqref{eq_final_duality}, $\langle \cdots \rangle$,
is taken for the Gaussian distribution
with mean $\overline{\bm{x}}^\mathrm{a}(\tau_{k-1})$ 
and covariance matrix $P^\mathrm{a}(\tau_{k-1})$.
The above procedure gives the mean $\overline{\bm{x}}^\mathrm{f}(\tau_k)$ 
and the covariance $\widehat{P}^\mathrm{f}(\tau_k)$ of the nonlinear systems at time $\tau_k$.
Note that the time-scaling variable $r_\mathrm{ts}$
must be selected adequately according to
the ratio between $\tau_{k} - \tau_{k-1}$ and $\Delta \tau$.
\item[4.] Assimilation step at $\tau_{k}$:\\
Calculate the Kalman gain
\begin{align}
\widehat{K}(\tau_k) = \widehat{P}^\mathrm{f}(\tau_k) H^\mathrm{T} 
\left( H \widehat{P}^\mathrm{f}(\tau_k) H^\mathrm{T} + R \right)^{-1},
\end{align}
where $H$ is the matrix assigned for the measurement and $R$ the variance of the measurement; 
see~Eq.~\eqref{eq_measurement}.
Update the following quantities:
\begin{align}
\overline{\bm{x}}^\mathrm{a}(\tau_k) = \overline{\bm{x}}^\mathrm{f}(\tau_k) + \widehat{K}(\tau_k) 
\left(
y(\tau_k) - H \overline{\bm{x}}^\mathrm{f}(\tau_k)
\right)
\end{align}
and
\begin{align}
P^\mathrm{a}(\tau_k) = 
(1 - \widehat{K}(\tau_k) H)
\widehat{P}^\mathrm{f}(\tau_k).
\end{align}
\end{enumerate}
Steps 3 and 4 are performed for each measurement time step.

\section{Results}

\begin{figure}
\includegraphics[width=90mm]{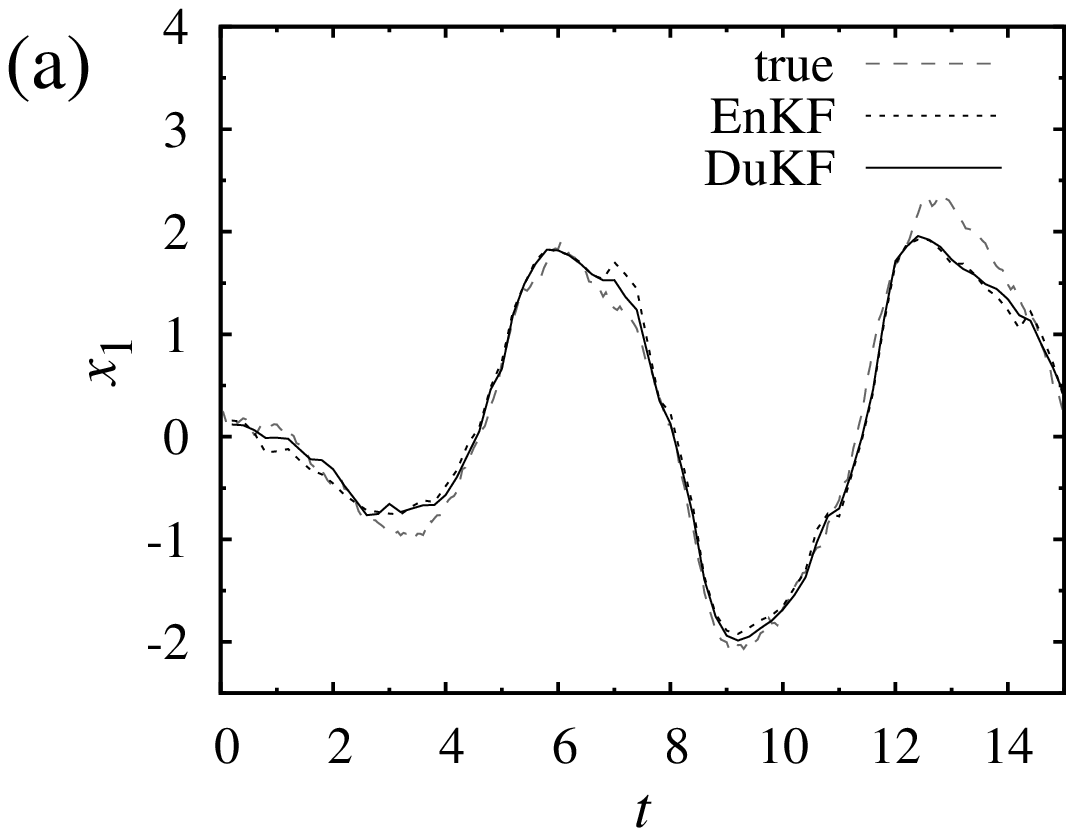}\\
\includegraphics[width=90mm]{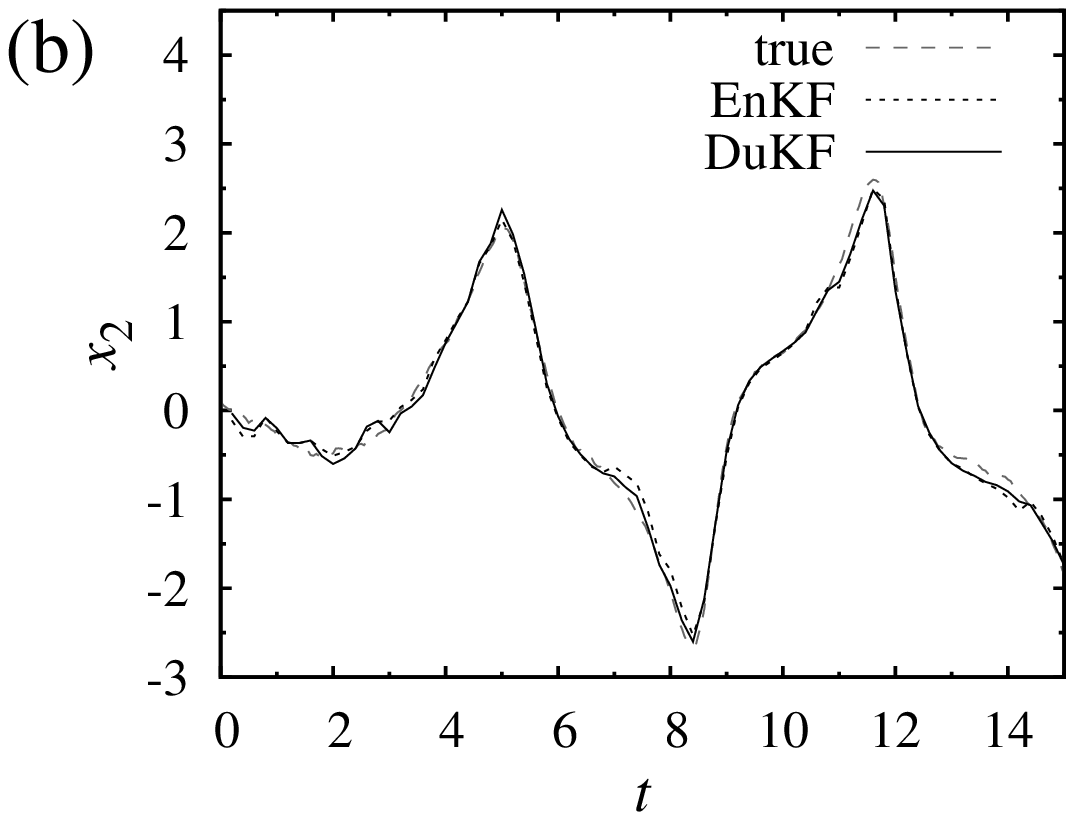}
\caption{
State estimates $x_1$ and $x_2$ for 
the stochastic differential equations~\eqref{eq_model};
(a) for the unobserved state variable $x_1$,
and (b) for the state variable $x_2$, which is observed with noise.
Note that $x_2$ is measured only at discrete times as shown in Fig.~\ref{fig_1}.
For the EnKF, we set $n = 10$ and $\Delta t = 10^{-4}$
in the Euler-Maruyama scheme.
}
\label{fig_2}
\end{figure}

In order to demonstrate the DuKF,
we employ the algorithm in Sec.~IV.E to the problem in Sec.~II.
For simplicity, we assume here that the parameters in the stochastic differential
equations, $Q$ and $R$, are previously known, as explained in Sec.~II.
We use the following parameters for the initial Gaussian distribution;
$\overline{\bm{x}}(0) = (0.1, 0.1)$,
$[P(0)]_{11} = 0.1$, $[P(0)]_{12} = [P(0)]_{21} = 0$, and $[P(0)]_{22} = 0.1$.
Firstly, for the DuKF, we performed the Monte Carlo simulations
for the dual birth-death process using the Gillespie algorithm \cite{Gillespie1977}.
For each initial condition, $10^{12}$ sample paths were generated,
and the Feynman-Kac term and the probability distribution were evaluated.

Figure~\ref{fig_2} shows the results of the state estimation.
Note that only the state variable $x_2$ is observed at discrete times,
as depicted in Fig.~\ref{fig_1}.
Although the observation contains the measurement noises, as seen in Fig.~\ref{fig_1},
both the EnKF and DuKF give adequate estimations.
Especially, the non-observed state $x_1$ can be estimated reasonably as shown in Fig.~\ref{fig_2}.
Note that the initial guesses for $\bm{x}$ are not far from 
the true values, and hence it is difficult to see the initial transient behavior
for both the EnKF and DuKF.
From Fig.~\ref{fig_2}, it may be difficult to judge
whether the DuKF gives better results than the EnKF or not,
but we confirmed that the DuKF gives a slightly better mean squared error.
If we want to obtain the similar mean squared error,
we need larger ensemble for the EnKF,
and the larger ensemble needs more computational time.
On the other hands, the forecast step in the DuKF 
does not need any Monte Carlo simulation,
and hence the DuKF works rapidly.
Actually, in order to deal with the current data,
the EnKF with ensemble size $n=10$ and original 
discretized time step for simulations ($10^{-4}$)
needs about $0.8$ seconds in a standard computer with Intel Core i5 processor
(2.2GHz).
In contrast, the DuKF needs less than $0.1$ seconds.
As shown below, the ensemble size $n=10$ is not enough large,
and if we use $n=1000$ ensembles,
$1000$ times costs of simulations are needed for the EnKF.
(Of course, the DuKF needs pre-calculations for the dual birth-death processes.
For $10^{12}$ sample paths, about one month calculations were needed,
but we can reuse the pre-calculation results repeatedly
in the actual filtering steps.)

\begin{figure}
\includegraphics[width=90mm]{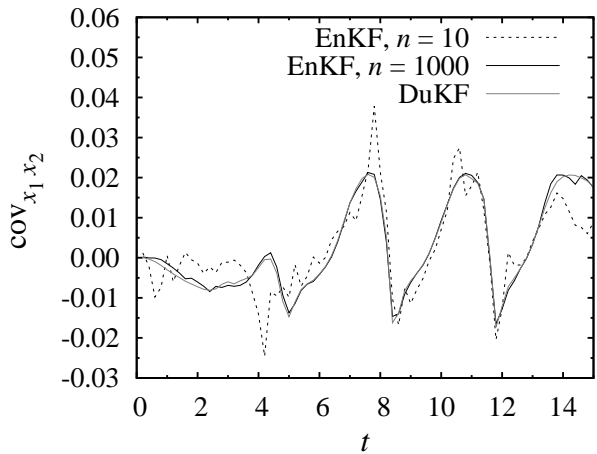}
\caption{
Covariance between $x_1$ and $x_2$, $[P(t)]_{12}$, calculated in the EnKF and DuKF.
The evaluated covariance is used to calculate the Kalman gain.
The evaluation in the EnKF approaches
to that in the DuKF when larger ensembles are used.
}
\label{fig_3}
\end{figure}

In order to see the difference between the EnKF and DuKF more explicitly,
we show the covariance $[P(t)]_{12}$ calculated in the EnKF and DuKF
in Fig.~\ref{fig_3}.
As shown in Fig.~\ref{fig_3},
the larger ensemble size in the EnKF gives
the similar covariance with that of the DuKF.
The covariance is used to evaluate the Kalman gain,
and hence more accurate covariance is necessary to have better estimations.
The ensemble size of order $10^3$ is needed
to obtain the similar covariance with that of the DuKF;
it is very time-consuming.
In addition, it would be needed to choose an adequate
time-interval $\Delta t$ for the simulation in the Euler-Maruyama scheme
for the stochastic differential equations;
the computational time and the precision of the estimations in EnKF
largely depend on $\Delta t$.
On the other hand, we do not need such time discretization
for the DuKF when the Gillespie algorithm is employed, as discussed before.

\section{Concluding remarks}

The duality relation between stochastic processes has been still developing,
and hence it is expected that new applications of the duality relations
give completely novel algorithms for various research fields.
We demonstrate one of the applications by using the estimation problem.
As a result,
pre-calculations for the dual birth-death processes
enable us to avoid the time-consuming Monte Carlo simulations
for each forecast step.

Of course, we do not claim that
the proposed DuKF is always superior to the EnKF for any estimation problem.
Actually, the implementation of the EnKF is simpler than that of the DuKF.
In addition, there could be more appropriate methods
for specific tasks (see, for example, \cite{Palatella2013}.)
In order to make the DuKF more practical,
it is needed to develop more efficient numerical methods
for the dual birth-death processes;
very high accuracy is necessary.
Actually, in preliminary works, the famous Lorenz system was investigated,
but the DuKF works well only for very short observation time intervals,
and then it was not practical.
If we have more accurate numerical results,
it would be possible to make the DuKF for the chaotic systems.
In order to overcome this problem,
it may be possible to employ simulations based on the important sampling methods
\cite{Asmussen_book,Landau_book}.
Note that it has not yet been clarified 
what factors determine the pre-calculation time;
the preliminary works for the Lorenz systems
could suggest that the chaotic behavior
is related to the demands for the high-accuracy of the pre-calculation,
but more detailed studies should be done in future
from both experimental and theoretical view points
for the practical usage of the duality relations.
We hope that the present work open up a new way of
the numerical applications of the duality concepts,
and further efficient numerical methods will be proposed in future.

\section*{Acknowledgement}

This work was supported in part by grant-in-aid for scientific research (No. 25870339)
from Japan Society for the Promotion of Science (JSPS).

\end{document}